# Corrections for the Geometric Distortion of the Tube Detectors on SANS Instruments at ORNL

Lilin He[1§]*, Changwoo Do[1§], Shuo Qian[1§], George D. Wignall[1], William T. Heller[1], Kenneth C. Littrell[2], Gregory S. Smith[1]

[1] *Biology and Soft Matter Division, Neutron Sciences Directorate, Oak Ridge National Laboratory, Oak Ridge, TN 37831, USA*
[2] *Chemistry and Engineering Materials Division, Neutron Sciences Directorate, Oak Ridge National Laboratory, Oak Ridge, TN 37831, USA*

[§]*L.H., C. D. and S. Q. contributed equally to this work*

***To whom correspondence should be addressed; Email:** *hel3@ornl.gov*

**Abstract** The small-angle neutron scattering instruments at the Oak Ridge National Laboratory's High Flux Isotope Reactor recently upgraded the area detectors from the large, single volume crossed-wire detectors originally installed to staggered arrays of linear position-sensitive detectors (LPSDs), based on the design used on the EQ-SANS instrument at ORNL's Spallation Neutron Source. The specific geometry of the LPSD array requires that approaches to data reduction traditionally employed be modified. Here, two methods for correcting the geometric distortion produced by the LPSD array are presented and compared. The first method applies a correction derived from a detector sensitivity measurement performed using the same configuration as the samples are measured. In the second method presented here, a solid angle correction derived for the LPSDs is applied to data collected in any instrument configuration during the data reduction process in conjunction with a detector sensitivity measurement collected at a sufficiently long camera length where the geometric distortions are negligible. Both methods produce consistent results and yield a maximum deviation of corrected data from isotropic scattering samples of less than 5% for momentum transfers up to a maximum of 0.8 Å$^{-1}$. The results are broadly applicable to any SANS instrument employing LPSD array detectors, which will be increasingly common as instruments having higher incident flux are constructed at various neutron scattering facilities around the world.

**Introduction**

With the ability to structurally characterize partially or completely disordered materials over length scales ranging from the sub-nanometer to a few hundred nanometers, small-angle neutron scattering (SANS) is widely used in many scientific disciplines including polymer science (Higgins & Benoit, 1994, He et al. 2012, Wignall & Melnichenko, 2005), molecular biology and biophysics (Martin et al., 2013, Heller & Baker, 2009, Svergun & Koch, 2003), metallurgy and condensed matter physics (Das et al., 2012, Zhang et al., 2011). Developments in SANS instrumentation, particularly detector technologies, have lagged behind those of its sister technique, small-angle x-ray scattering (SAXS), because of the far slower progress in neutron source development, as well as and the lower abundance of neutron sources and beamlines relative to synchrotrons. The single volume, crossed-wire gas detectors traditionally used suffer from count-rate limitations that hinder SANS instruments on many of the brightest neutron sources. For example, detector limitations have prevented the application of SANS to sub-minute studies of kinetic processes in strongly scattering samples that are readily addressed with SAXS (Svergun & Koch. 2003).

In order to mitigate this count-rate limitation on the GP-SANS (Wignall et al, 2012) and Bio-SANS (Lynn et al, 2006) instruments at the High Flux Isotope Reactor (HFIR) of Oak Ridge National Laboratory (ORNL), the crossed-wire position-sensitive detectors have been replaced by linear position-sensitive detectors (LPSD) arrays (Berry et al., 2012). Other neutron scattering facilities have also implemented LPSD detector arrays for low-q and intermediate-q instruments (Otomo T. et al., 1999, Thiyagarajan, P. et al., 1998). To our knowledge, the D22 and D33 at the Laue-Langevin (Grenoble, France) were the first two SANS instruments to be equipped with such a large multitube detector array that are still in service. (http://www.ill.eu/instruments-support/instruments-groups/instruments/). The LPSD arrays installed on the SANS instruments located at HFIR utilize the design implemented on the EQ-SANS instrument of the Spallation Neutron Source (SNS) (Zhao et al., 2010). The 1 m long LPSDs (GE Measurement & Control; Twinsburg, OH, USA) are made of thin-walled (0.25mm thick) stainless steel tubing (Berry et al., 2012). For the instruments at HFIR, each tube is filled with 10 atm $^3$He gas for neutron conversion and contains a quench gas that is a mixture of mostly argon with a small amount of $CO_2$, giving a total pressure of 10.9 atm, while the tubes used on EQ-SANS have 20 atm of $^3$He and 2.1 atm of the quench gas to improve the detection efficiency for the shorter wavelength neutrons commonly employed on the instrument (Zhao et al., 2010). Each tube contains a coaxial electrode and operates in proportional mode.

For ease of assembly, handling, and testing, LPSDs are grouped into modules of 8 detectors, called an "8-pack", using electronics developed at ORNL. In contrast to the single detection plane LPSD array in use on the D22 and D33 SANS instruments of ILL, each 8-pack module consists of front and back rows of four offset tubes, shown schematically in Figure 1, that are affixed in an extruded aluminium frame that keeps the tubes straight when mounted vertically. The use of offset front and back LPSD planes with a gap between tubes less than the tube width presents a contiguous gas volume to neutrons incident on the 8-pack (Zhao et al., 2010). The entire detector array contains 24 8-packs (192 tubes) mounted side by side onto a common frame. The modular design makes it possible to turn off any module or tube that loses function during a neutron production cycle without shutting down the instrument to repair the whole array, which both simplifies maintenance and minimizes the loss of precious neutron beam time.

The new detectors at HFIR entered operation in the ORNL user program in 2011 and have performed robustly and reliably at count rates of up to ~1000 counts-per-second (cps) per pixel and $10^6$ cps globally (Berry et al., 2012). These rates are orders of magnitude higher than could be sustained on the original crossed-wire gas detectors, enabling users to more efficiently study strongly scattering materials that required beam attenuation when measured on the original detectors. By virtue of being able to withstand much higher count rates, the new detectors enable time-resolved studies on sub-minute time scales. Further, samples having a large incoherent scattering background, i.e. proteins at low concentration in $H_2O$, and samples for contrast-matching studies of small domains in large multi-domain molecules (Crawford et al., 2013, Boukhalfa et al., 2013) benefit from reduced intensity variance that results from the much higher count rates accepted by the new detectors.

While the new LPSD array detectors have improved the performance of the HFIR SANS instruments significantly, they produce different geometric distortions in the data collected than those observed with crossed-wire detectors, including those that are particularly strong at short sample-to-detector distances that give rise to large scattering angles. These effects arise from both the shadowing of the back plane of tubes by those in front and a different solid angle correction necessitated by the use of LPSDs. Here, we present two methods for correcting the distortions that arise in the data and validate them by performing a series of measurements on well-characterized samples. The results demonstrate that both methods can serve to correct SANS data from the LPSD arrays, provided that certain limitations of their applicability are respected.

**Experimental**

A series of samples known to produce isotropic scattering were measured: 2.3 mm thick single-crystal vanadium, 1 mm of $D_2O$, 1 mm of $H_2O$, 1.4 mm thick poly(methyl methacrylate) (PMMA). As discussed by Wignall (Wignall, 2011), the spatial variation of the detector counting efficiency has hitherto been measured via the scattering from these standards. While the multiple scattering processes in such materials are not well-characterized, the scattering of neutrons by predominantly protonated materials is isotropic. Thus, the variation in the measured signal is proportional to the detector efficiency to a good approximation and may be used in the data reduction process to correct for this effect on a pixel-by-pixel basis. When corrected via an H$_2$O detector efficiency measurement, the scattering from such materials is independent of the scattering angle to an excellent approximation.

The measurements were performed on GP-SANS instrument at HFIR (Wignall et al., 2012). The neutron wavelength, $\lambda$, was 4.7 Å, while the wavelength spread, $\Delta\lambda/\lambda$, was set to 0.13 using the velocity selector. A 40 mm diameter source aperture and a 12mm diameter sample aperture were used to collimate the incident beam. A sample-to-detector distance (SDD) of 1.1 m was used with the centre of the detector offset from the beam position by 40 cm to reach the maximum momentum transfer $Q$ ( $Q = 4\pi \sin(\theta)/\lambda$ where $2\theta$ is the scattering angle and $\lambda$ is the neutron wavelength) of the instrument. The data acquisition time for each sample was approximately 10 minutes. In addition, a 33 mm thick PMMA block was measured for 5 hours at an intermediate SDD of 6.8m for use as a detector efficiency measurement in the analytical data correction. At this SDD, both the solid angle correction over the detector and the angular dependence of the sample transmission are negligible. A 10-minute measurement of single crystal vanadium in the 1.1 m configuration described above was used as the sensitivity measurement for the in-situ correction approach. Prior to being azimuthally averaged to produce a 1D profile *I(Q)*, the 2-dimensional raw counts were corrected for air scattering, the angular dependence of the transmission and the dark current resulting from the ambient radiation background and electronic noise (Wignall & Bates, 1987).

**Results and Discussion**

**1. Shadow analysis of tubes**

**1.1 Tubes in the same plane**

The detector is composed of two planes, each of which contains 96 tubes as previously mentioned. Figure 2 shows the top-down view of a pair of adjacent tubes in a single plane. The line labelled ACEC' is tangent to both tubes at the points C and C'. It is clear that if the SDD is less than a certain distance, a tube will be partially shadowed by the one immediately adjacent to it that is closer to the direct beam position. This distance is determined by the distance between the middle point of the two neighbouring tubes (OO'); the distance from the beam centre to the midpoint between the pair of tubes (DE); and the radius of the tube (CO). The minimum SDD free of tube-tube shadowing can be calculated based on the geometry of the farthest tube from the beam centre in the rear plane, which is given by

$$SDD = \frac{CO \cdot DE}{\sqrt{\left(\frac{OO'}{2}\right)^2 - CO^2}}. \qquad (1)$$

For the GP-SANS and Bio-SANS detectors, CO=7.94/2=3.97mm, OO'=11mm (Berry et al., 2012), and the maximum accessible value of DE is 900mm. In the case of the EQ-SANS detector, OO' = 8.2 mm (Zhao et al., 2010). For the HFIR instruments, equation (1) gives a shadow-free minimum SDD of 938.67mm ($2\theta > 43.8°$) based on the equation (1), which is less than the minimum sample-detector distance possible on either GP-SANS or Bio-SANS instruments. The detector cannot be offset on the EQ-SANS and the minimum SDD is 1.3 m, giving a smaller maximum scattering angle ($2\theta > 21.0°$) than the HFIR instruments. Therefore, same-plane shadowing is not an issue for any of the ORNL SANS instruments.

**1.2 Shadow of front tubes on the back ones**

In contrast to tubes in the same plane, the staggered arrangement of tubes in different planes means that the tubes in the back plane will always be shadowed by those in the front plane in an instrument configuration-dependent manner. The shadowing can be clearly seen in Figure 3 in the data collected from the single crystal vanadium in the 1.1 m configuration described above. As expected, the count rate of the tubes in both planes decreases with increasing distance from the beam position due to the reduction in solid angle viewed by each pixel with increasing scattering angle. Furthermore, the shadowing of the back tubes by those in the front plane causes a faster rate of count rate decay in the back plane of tubes. This trend reverses after tube #156 (counting from the left in the image), where the most significant shadowing is expected for this instrument configuration.

The illumination factor of each pixel on the back panel can be calculated based on the viewing angle through which each pixel detects neutrons. Four specific geometric shadowing conditions exist, which are described below and illustrated in Figure 4.

(I) When the scattering angle is small (see (a) in Fig. 4), ∠CAD < ∠B'AD < ∠FAD <

∠C'AD, tube C will be partly blocked by both tube B and tube F. The angle for tube C to detect neutrons is determined by

$$\phi = \angle FAD - \angle B'AD$$

(II) When ∠CAD < ∠B'AD < ∠C'AD < ∠FAD, tube C will be partly blocked by tube B (see (b)), the angle through which tube C can detect neutrons is determined by

$$\phi = \angle C'AD - \angle B'AD$$

(III) When ∠BAD < ∠CAD < ∠C'AD < ∠B'AD, tube C will be totally blocked by tube B (see (c) in Figure 4), the angle through that tube C can detect neutrons becomes zero. This also can be regarded as the special case of case II where the angle through which tube C can detect neutrons is zero.

(IV) When ∠G'AD < ∠CAD < ∠BAD < ∠C'AD, tube C will be blocked by tube B and neutrons will fall on the left side of the tube (see (d)), the angle is given by

$$\phi = \angle BAD - \angle CAD$$

The illumination factor of tube C should be calculated by

$$f = \phi/\varphi$$

where φ or ∠CAC' is the total angle for the tube C to see neutrons without any shadow of the front tubes at a configuration.

All pixels on the same tube share the same shadow factor resulting from the arrangement of the tubes in the array. Therefore, one only needs to determine the scattering angles of those tangential points (Figure 5) located in the horizontal plane containing the incident beam to calculate the shadow factor of each back-plane tube: i.e, for the tangential point B in Figure 5, the scattering angle is calculated according

to:

$$\angle BAD = \arcsin\left(\frac{X_B}{\sqrt{X_B^2+SDD^2}}\right) - \arcsin\left(\frac{r}{\sqrt{X_B^2+SDD^2}}\right) \quad (2)$$

where $X_B$ is the x coordinate of the tube B and r is the radius of the tube. The scattering angle of the tangential point B', similarly, is determined by

$$\angle B'AD = \arcsin\left(\frac{X_B}{\sqrt{X_B^2+SDD^2}}\right) + \arcsin\left(\frac{r}{\sqrt{X_B^2+SDD^2}}\right) \quad (3)$$

The red dotted line in Figure 6 corresponds to the illumination factors calculated for the tubes from an instrument configuration where the incident beam is offset to the largest degree and the sample is placed at the shortest detector position. It should be noted that the centre of the EQ-SANS detector cannot be offset significantly from the incident beam, but the detector can be set at a similarly short sample-to-detector distance (1.1 m on the HFIR SANS compared to 1.3 m on the EQ-SANS).

A comparison of the calculated illumination factors in terms of geometries with the experimental illumination factors of the back plane tubes from single crystal vanadium is shown in Figure 6. The experimental factors were obtained by using the count of each back tube divided by that of its immediate next one in the front plane after the corrections for background, transmission and solid angle. The experimental factor deviates from the predicted value with as the amount of shadowing increases, which indicates that other processes take place in the detector beyond those dictated by geometry. Specifically, the transmission of neutrons through the front plane of tubes and the subsequent detection of a comparable fraction of these transmitted neutrons in the back plane of tubes causes the deviation. While the high pressure of $^3$He (10 atm in the HFIR instruments and 20 atm in the EQ-SANS detector) is anticipated to provide a very high efficiency for the wavelengths that the instruments were designed for (> 4.5 Å at HFIR and > 2.0 Å at the EQ-SANS), neutron scattering processes in most materials produce a considerable fraction of inelastic scattering events, with the most common process being an energy gain (Do et al., 2014). These neutrons have a considerably shorter wavelength (1 Å or 2 Å corresponding to the thermal energy) than the incident beam, which in turn means that their detection efficiency in the LPSDs is lower than the wavelengths for which they were designed. Such effects exist in single plane LPSD arrays, such as the D22 and D33 detectors, and in single volume, crossed-wire detectors, but they are not observed because there is no second plane of detection. In order to correct for this effect, the transmissions of all the tubes and the shadow factor of each pixel in the detector must be determined for each different configuration used. Although the latter can be calculated from the viewing angle through which each pixel detects neutrons, measuring the transmission of each tube is rather challenging because the transmission depends not only on the path length through the tubes of the scattered neutrons, but also on the scattered neutron wavelength, which is altered by the material-dependent inelastic scattering processes in samples. As a result, developing an analytical correction for the detection efficiency of the tubes that can be applied to data collected in both planes of the detector is impractical.

## 2. Strategies for performing high-Q corrections for tube detectors

Data from crossed-wire detectors have traditionally been divided by $\cos^3(\theta)$ to correct for the variation in the solid angle subtended by different pixels present in a planar geometry ( Glinka, C., et al., 1998, Grillo, 2012, Wignall et al., 2013), As noted in Section 1.2, this correction is not suitable for the LPSD arrays on the SANS instruments at ORNL, or for LPSD arrays in general, but is suitable for single volume, crossed-wire detectors in samples displaying little inelastic scattering. Application of this correction to data collected using the ORNL LPSD design results in a substantial overcorrection observed as an "upturn" in the high-Q region of the data for the isotropically-scattering standard samples mentioned above. The upturns in data collected on the HFIR SANS instruments are ~17% at $Q \sim 0.8$ Å$^{-1}$. Here, we present two possible approaches for correcting for the geometric distortions.

## 2.1 Data correction using a detector sensitivity measured at the same configuration

As already shown in the section 1.2, the tubes in the rear plane are shadowed by those in the front one. An approach for recovering the data collected by the back tubes is to employ an "in-situ" correction during the data reduction. The word "in-situ" here means that a detector sensitivity measurement is performed using the same instrument configuration as the sample measurement. The mathematical solid angle corrections are not applied, but the correction for the angular dependence of the sample transmission is applied to account for the fact that the sample and standards have different transmissions. A single crystal vanadium sample was selected for the in-situ sensitivity due to its high fraction of elastic scattering relative to hydrogenous materials (Ghosh and Rennie, 1999, Belmabkhout and Sayari, 2009). The 2D data of the sample was normalized to that of vanadium in terms of the equation

$$I'(x,y) = \frac{\left(\frac{I(x,y)}{T} - \frac{I_{background}(x,y)}{T_{background}}\right)\left(\frac{d_{Vanadium}}{d}\right)}{\left(\frac{I_{Vanadium}(x,y)}{T_{Vanadium}} - \frac{I_{background}(x,y)}{T_{background}}\right)} \quad (4)$$

where d represents the thickness of the sample or vanadium. The background in this case arises from air scattering. In addition, the transmissions T for both sample and vanadium are corrected for the angular dependence using their zero-angle transmission, T(0), by means of equation 5 (Lindner et al., 2000, Hamilton, W. A., 2007).

$$T(2\theta) = T(0)^{[1+\sec(2\theta)]/2} \quad (5)$$

The corrected 2D data for 1mm $D_2O$ is shown in Figure 7 (left). Looking along a horizontal line through the beam centre (Figure 7, right) shows that the corrected data from the front tubes are essentially flat. In contrast, the corrected data from the back tubes show a broad peak that begins near tube #75 tube and is centred around tube #156. A deviation from isotropic scattering of ~20% at Q ~ 0.8 Å$^{-1}$ was found when using all tubes in the front and back planes. As mentioned in section 1.2, the incoherently scattered neutrons are thermalized during the scattering events and produces neutrons of thermal energy (~20meV) that have a shorter wavelength (~2 Å) than the incident neutron wavelength (4.6 Å), leading to higher transmission through the front tubes. (Do et al., 2014) Therefore, the high-q upturn can be attributed to transmitted neutrons being detected in the back plane of tubes in addition to those directly incident on these tubes. As a result, some or all of the tubes in the back plane must be masked out in the data reduction for best data reduction results for short SDD configurations. Masking out the most strongly impacted detector tubes produces corrected data from $D_2O$ that is isotropic in Q out to 0.6 $\text{Å}^{-1}$, which is generally sufficient for studies of soft matter and biology (shown in Figure 8).

The performance of the in-situ sensitivity data reduction when the whole back plane of tubes was masked was evaluated through a series of measurements with an incident wavelength of 4.7 Å at the closest SDD (~ 1 m). The single crystal vanadium again was chosen for the sensitivity measurement and the results are presented in Figure 9. By using these procedures, the Q-dependence of the data from 1 mm $D_2O$ is reduced, varying less than 2% over the detector. The profiles of PMMA and $H_2O$, which both display more incoherent scattering than $D_2O$, show a negative departure (up to 5%) at high-Q (see Figure 9). Further improvements would require corrections for the dependencies of detection efficiency on wavelength and neutron trajectories through the tubes (see Supplementary Information). As noted, these corrections would ultimately be sample-dependent and, therefore, impractical to implement.

**2.2 Data correction using a modified solid angle correction derived for tube detectors**

The "in-situ" method is an effective means of correcting SANS data, even at high Q, as shown above and previously (Wignall, 2011). However, the measurement of the detector sensitivity for each different configuration at which samples are measured is not always convenient, especially when complex and unconventional sample environments are used or when studying weakly scattering samples. Thus, a data correction method must be developed that can use a sensitivity measurement collected with an instrument configuration that requires minimal solid angle corrections, such as at a SDD ≥ 6.8 m with no offset of the centre of the detector from the position of the primary beam.

The traditional solid angle correction for flat detectors is given by:

$$\Delta\Omega(2\theta) = \frac{p_x p_y \cos^3(2\theta)}{SDD^2} \qquad (6)$$

where $p_x$ and $p_y$ are the $x$ and $y$ dimensions of the pixel, respectively, $2\theta$ is the scattering angle. The cylindrical geometry of a pixel in the LPSD, shown in Figure 10, causes the solid angle subtended by a pixel to change differently with the scattering angle along the horizontal direction than it does along the vertical direction. According to the definition of solid angle, we have

$$\Delta\Omega(2\theta) = \frac{p_x p_y \cos(\alpha)}{D^2(2\theta)} \qquad (7)$$

where α is ∠BAC in Figure 10, which is in the plane that is normal to the horizontal plane ADC and parallel to the tubes; $D(2\theta)$ is the distance from the source to pixel with a scattering angle of 2θ

since $$D(2\theta) = SDD/\cos(2\theta)$$

Therefore,

$$\Delta\Omega(2\theta) = \frac{p_x p_y \cos(\alpha)\cos^2(2\theta)}{SDD^2} \qquad (8)$$

Note that $p_x$ can be approximated by the tube diameter because the tube diameter is small compared to the SDD. For the pixels located on the central tube (along DE in Figure 10), α =2θ, and the equation (8) becomes the traditional solid angle correction. For the pixels on the horizontal line of DC, α=0, the equation (8) turns into

$$\Delta\Omega(2\theta) = \frac{p_x p_y \cos^2(2\theta)}{SDD^2} \quad (9)$$

The change from $cos^3(2\theta)$ to $cos^2(2\theta)$ is the reason that the decrease in solid angle along the horizontal direction is slower than that along the vertical direction. The 1D profiles of vanadium, $D_2O$, $H_2O$ and PMMA reduced using Equation (9) and a detector sensitivity measurement collected at 6.8 m are plotted in Figure 11. This new formula eliminates the overcorrection at high scattering angles (2θ > 10°) that is produced by the old solid angle correction. Generally a maximum deviation of less than 2% for both vanadium and $D_2O$ up to 0.8 Å$^{-1}$ is achieved. However, a deviation of 5% for PMMA and $H_2O$ is observed, which is again attributed to the large amount of hydrogen that produces inelastic scattering in these samples and the impact of the neutron path-length on the detection efficiency of the tubes. It should also be noted that the overcorrection using the traditional solid angle correction affects not only the "flat" patterns, but also the sharp features at large Q domain (larger than Q > 0.5Å$^{-1}$), which could result in misinterpretation of the data with increasing scattering angle (see S.2 of the Supporting Information).

**Summary**

Two different approaches for correcting the geometric distortions observed with the new detectors installed on the SANS instruments of ORNL's HFIR, the GP-SANS and the Bio-SANS, have been developed, compared and demonstrated to be effective. Generally, a maximum variation over the detector of less than 5% can be achieved for the samples that exhibit isotropic scattering patterns. The improved performance and the consistency between the results of the two approaches reinforce their validity. However, the in-situ sensitivity method requires a different sensitivity run for each scattering geometry employed, thereby requiring users of the instrument to dedicate awarded beam time to these calibration measurements. In contrast, the modified solid angle correction can be applied to any data collected with the instruments using a sensitivity collected during normal calibration activities performed regularly by the instrument staff, making this the preferable method for correcting the SANS data for routine use. Furthermore, the analysis of the shadowing of the back plane of tubes by the front plane shows that the back-plane data can be used in many cases, but that the use of extremely short SDD instrument configurations dictates that the back panel of tubes must be wholly or partially discarded in most cases to provide the best results without the need of further corrections. For the future instruments with LPSD-based detector banks, detectors used for collecting high scattering angle data ($2\theta > 25°$) should use a single plane of LPSDs to avoid tube shadowing. However, for low and intermediate scattering angles ($2\theta < 25°$), the use of a staggered array affords a continuous gas volume that maximizes neutron detection efficiency and improves spatial resolution of the detectors to less than the tube diameter. The work presented here shows that it is possible to adequately correct the data from such detectors, making their benefits outweigh the costs of the increased data complexity that results from the design.

## Acknowledgements

The research at Oak Ridge National Laboratory was sponsored by the Scientific User Facilities Division, Office of Basic Energy Sciences, U.S. Department of Energy. S.Q. acknowledges the support of the Office of Biological and Environmental Research of the United States Department of Energy through the Center for Structural Molecular Biology at Oak Ridge National Laboratory. The authors thank Yuri B. Melnichenko for providing single crystal vanadium sample and also thank Yuri B. Melnichenko, Volker S. Urban, Sai Venkatesh Pingali and Christopher Stanley for their technical discussions and Kevin D. Berry for information about the detection efficiency of the LPSDs.

## References

Belmabkhout, Y., and Sayari, A. (2009). Chem. Eng. Sci., 64, 3729-3735.

Berry, K. D., Bailey, K. M., Beal, J., Diawara, Y., Funk, L., Hicks, J. S., Jones, A. B., Littrell, K. C., Pingali, S. V., Summers, P. R., Urban, V. S., Vandergriff, D. H., Johnson, N. H., Bradley, B. J. (2012). Nuclear Instruments and Methods A, 693, 179-185.

Boukhalfa, S., He, L., Melnichenko, Y. B., Yushin, G. (2013). Angewandte Chemie

International Edition 52, 4618-4622.

Crawford, M.K., Smalley, R.J., Cohen, G., Hogan, B., Wood, B., Kumar, S.K., Melnichenko,Y.B., He, L., Guise, W., Hammouda, B. (2013) Phys. Rev. Lett. 110, 196001.

Das, P., Rastovski C., O'Brien T. R., Schlesinger K. J., Dewhurst C. D., DeBeer-Schmitt L., Zhigadlo N. D., Karpinski J., Eskildsen M. R. (2012). Phys. Rev. Lett. 108, 167001.

Do C., Heller W. T., Stanley, C., Gallmeier, F., Doucet, M., Smith, G.S. Nucl. Instr. Meth. Phys. Res. A 2014, 737, 42-46.

Ghosh, R.E., Rennie, A.R., (1999). J. Appl. Cryst. 32, 1157-1163.

Glinka, C.J., Barker, J,G., Hammouda, B., Kruger, S., Moyer, J. J. & Orts, W.J. (1998). J. Appl. Cryst. 31, 430-445.

Hamilton, W. A. (2007) Personal communication.

He, L., Cheng G., Melnichenko Y. B. (2012). Phys. Rev. Lett. 109, 067801.

Heller, W. T. and Baker, G. A. (2009). "Visualizing Structures of Biological Macromolecules with Small-angle Neutron Scattering and Modeling," in Neutron Imaging and Applications, Springer, New York, NY, pp. 289-304.

Higgins, J. S. & Benoit, H. C. (1994). Polymers and Neutron Scattering, 1st ed Oxford University Pres.

Lindner, P. Leclercq, F., Damay, P. (2000), Physica B., 291, 152-158

Lynn, G.W., Heller, W. T., Urban, V., Wignall, G.D., Weiss, W., Myles D. A. A.

(2006). Physica B: Condensed Matter, 385-386, 880-882.

Martin, S. L., He L., Meilleur F., Guenther R. H., Sit T. L., Lommel S. A., Heller W. T. (2013). Archives of Virology 158, 1661-1669.

Otomo, T., Furusaka, M., Satoh., S., Itoh, S., Adachi, T., Shimizu, S., Takeda, M. (1999). Journal of Physics and Chemistry of Solids 60, 1579-1582.

Svergun, D. S. and Koch, M. H. .J M. (2003). Rep.Prog.Phys. 66, 1735-1782.

Thiyagarajan, P; Urban, V; Littrell, K; Ku, C; Wozniak, DG; Belch, H; Vitt, R; Toeller, J; Leach, D; Haumann, JR; Ostrowski, GE: Donley, LI; Hammonds, J; Carpenter, JM; Crawford, RK, PROCEEDINGS OF ICANS-XIV June 14-19, 1998. Utica, IL, USA.

Wignall, G. D., Melnichenko Y. B. (2005). Rep. Prog. Phys. 68, 1761-1810.

Wignall, G. D., Littrell, K.C., Heller, W.T., Melnichenko, Y.B., Bailey, K.M., Lynn, G.W., Myles, D.A., Urban, V.S., Buchanan, M.V., Selby, D.L., Butler, P.D. (2012). J. Appl. Cryst., 45, 990-998.

Wignall, G. D.; Bates, F. S. (1987). J. Appl. Cryst. 20, 28.

Wignall, G. D. (2011). Neutrons in Soft Matter, edited by T. Imae, T. Kanaya, M. Furusaka & N. Torikai, PP. 285-309. Hoboken: John Wiley and Sons.

Wignall, G.D., Melnichenko Y. B., He L., Littrell Ken (2013). Notes on the Collection and Initial Analysis of Cold Neutron SANS Data on the ORNL GP-SANS facility.

Zhang, Z.W., Liu, C.T., Wang, X.L., Littrell, K.C., Miller, M.K., An, K., Chin, B. A. (2011). Phys. Rev. B. 84, 174114.

Zhao, J.K., Gao, C.Y., Liu, D., (2010), J. Appl. Cryst. 43, 1068-1077.

**Figure Captions**

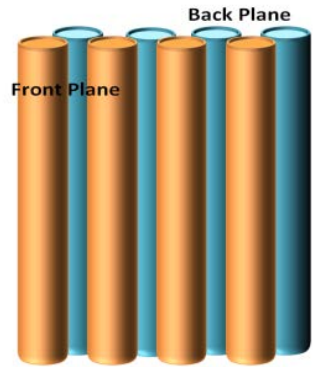

Fig. 1 Schematic picture of an 8-pack module showing the arrangement of the front and back planes of LPSDs.

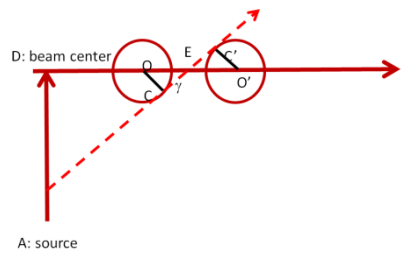

Fig. 2 Geometry of shadow analysis for the tubes in the same plane.

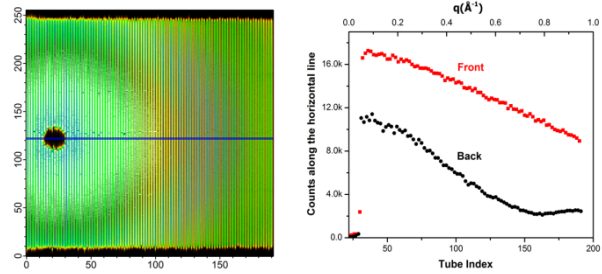

Fig. 3 Raw image of 2.3mm single crystal vanadium collected at 1m SDD (left), 7guides and 4.7Å. Horizontal cross section of the 2D image (right). Note that the x axis of the right graph starts from the beam centre instead of the leftmost tube of the detector.

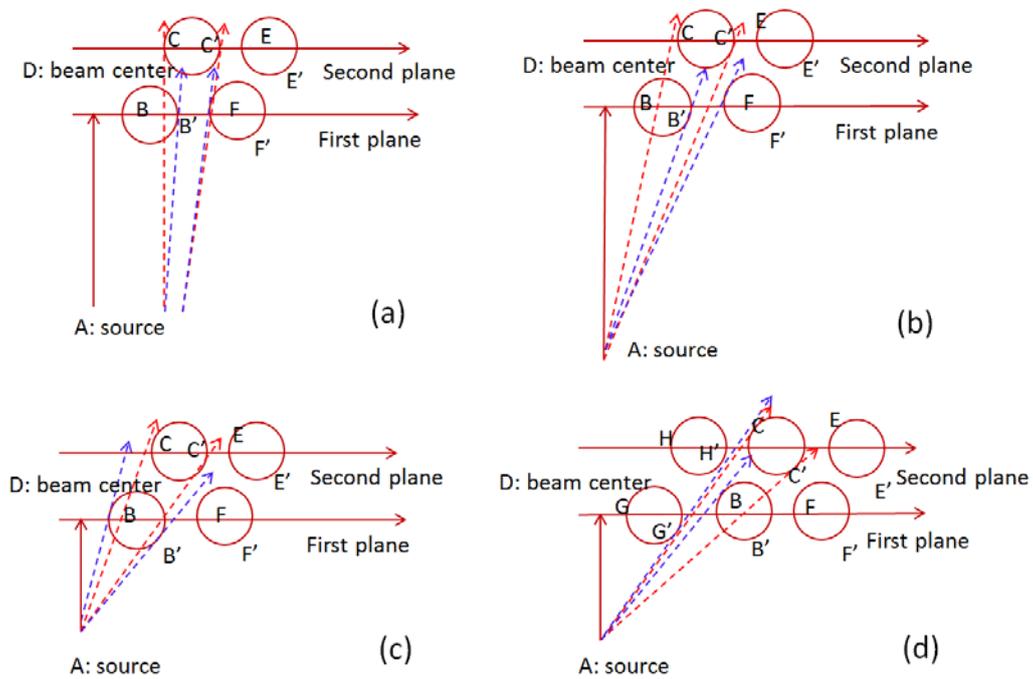

Fig. 4 Viewing angles of back tubes at different locations. Four cases need to be considered: Tube C is partly blocked by tube B and F (a), blocked by tube B (b), totally blocked by tube B (c) and partly blocked by tube B (d). Since all pixels on the same tube share the same shadow factor, only the scattering angles of those tangential points in the horizontal plane are calculated.

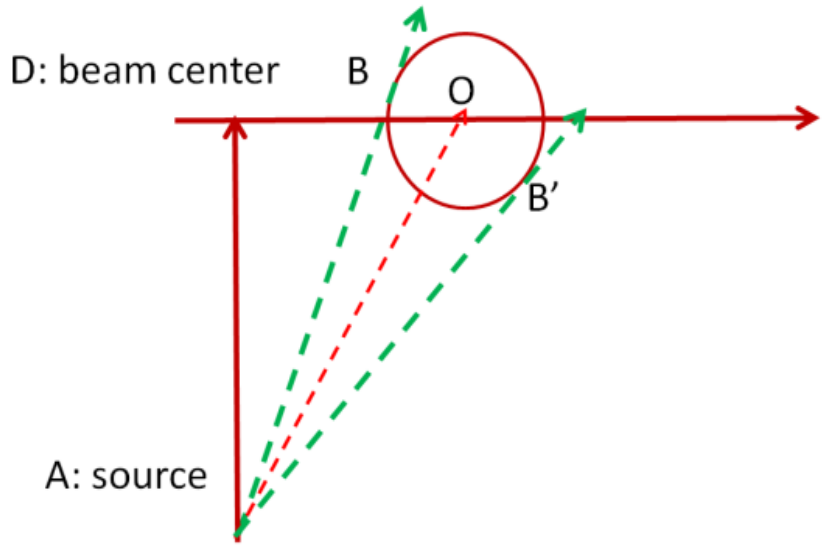

Fig. 5 Geometry for the scattering angles of the tangential points of the tubes in the horizontal plane.

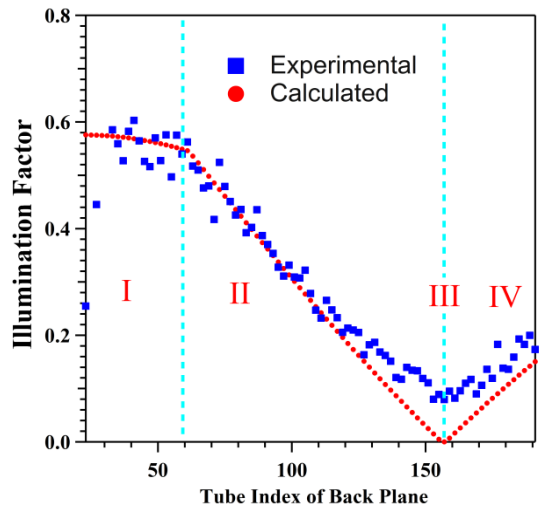

Fig. 6 Calculated and experimental illumination factor as a function of tube position for back-plane tubes (SDD=1.1m, detector offset = 400mm). The displayed tube index starts from the beam centre at tube #22 rather than the leftmost one. One should notice that the condition III in the main text refers to the single point at the minimum of the curve.

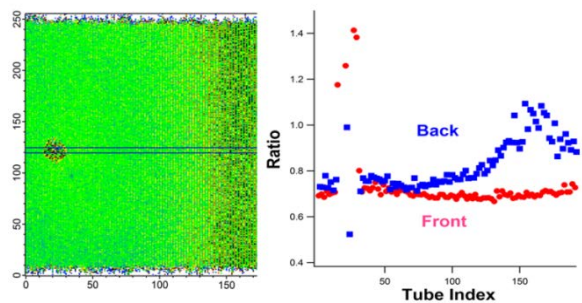

Fig. 7 Ratio of $D_2O$ data to vanadium data, collected at 1m SDD, 7guides and 4.7Å (left). Horizontal cross section of the 2D image (right). The "in-situ" correction method breaks down if all the back tubes are used due to a larger number of inelastic scattering events occurred in $D_2O$ than that in vanadium.

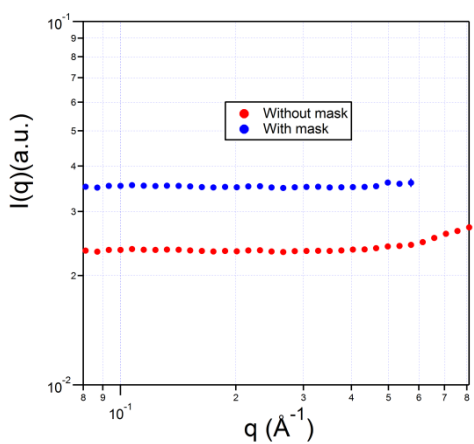

Fig. 8 SANS profiles of 1mm $D_2O$ corrected using vanadium measured at an exactly the same sample-detector distance as the sample. The data collected by both planes are used. A 20% positive deviation at Q ~ 0.8 Å$^{-1}$ is observed (red line) if all the tubes are used.    The deviation is minimized (less than 2%) if the tubes from #75 to #192 are masked (blue line).

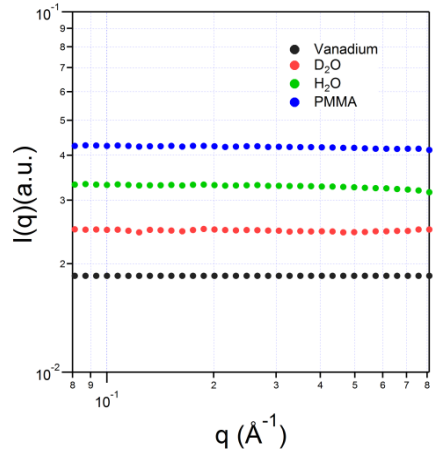

Fig. 9 SANS profiles of vanadium, $D_2O$, $H_2O$ and PMMA corrected with back tubes being masked using "in-situ" efficiency measured at an exactly the same sample-detector distance as the samples.

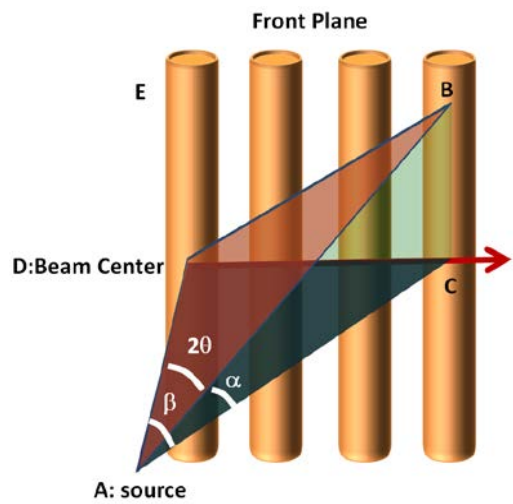

Fig. 10 Geometry of the front panel of tube detectors. A neutron strikes a tube at point B and its projection in the horizontal plane is C. The angle α is ∠BAC and angle β is ∠DAC, 2θ is the scattering angle ∠BAD of the pixel at point B. One may notice $\cos(2\theta) = \cos(\alpha)\cos(\beta)$

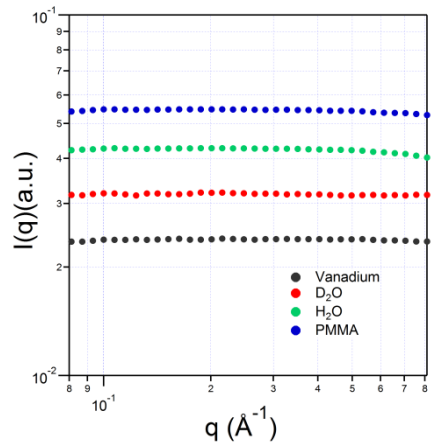

Fig. 11 SANS profiles of vanadium, $D_2O$, $H_2O$ and PMMA corrected using modified solid angle correction (Eq. 4) for tube detectors with the back tubes being masked.